# Pressure Study of Superconductivity and Magnetism in Pure and Rh-Doped $RuSr_2GdCu_2O_8$ Materials

M. Steiger,[a] C. Kongmark,[a,b] F. Rueckert,[a] L. Harding,[a]
and M. S. Torikachvili [a]

[a]*Department of Physics, San Diego State University, San Diego, CA 92182-1233, USA*
[b]*Department of Chemistry, Université des Sciences et Thechnologies de Lille (Lille1),*
*Velleneuve d'Ascq Cedex, 59652, France*

**Abstract**

A hydrostatic pressure study was made on pure and Rh-doped specimens of the superconducting ferromagnetic compounds $Ru_{1-x}Rh_xSr_2GdCu_2O_8$ ($x$ = 0-0.15) by means of measurement of electrical resistivity vs temperature, in pressures up to 2 GPa. Partial substitution of Rh for Ru decreases the magnetization of the material, lowers both the magnetic ordering temperature $T_m$, and the superconducting transition temperature $T_c$, and promotes granularity. The effect of pressure for all compositions is an increase in both the intra- and inter-granular superconductivity transition temperatures, $T_c$ and $T_p$ respectively, as well as $T_m$. The rate of change of each transition temperature with pressure first drops for Rh concentrations near 5%, increasing latter for higher concentrations. While the rate of increase of $T_c$ with pressure for all compositions is 2-3 times lower than in YBCO materials, the simultaneous increase of $T_c$ and $T_m$ with pressure could support the notion of competition between superconductivity and ferromagnetism in these materials. The effect of pressure on the weak-links was a significant improvement of inter-granular connectivity.

**Introduction**

The ruthenocuprates of general composition $RuSr_2LnCu_2O_8$ (Ru-1212; $Ln$ = Eu, Sm, Gd) were first synthesized in 1995 [1]. The Ru-1212-Gd compound exhibits long range magnetic ordering (MO) below the ordering temperature $T_m \approx 135$ K and superconductivity (SC) below the transition temperature $T_c \approx 35$ K [2], resulting in coexistence of SC and MO below $T_c$. The MO that develops at $T_m \approx 135$ K is due to the Ru sublattice. Measurements of magnetic susceptibility and Mössbauer spectroscopy in the related material $RuR_{1.4}Ce_{0.6}Sr_2Cu_2O_{10-\delta}$ (Ru-1222; $R$ = Eu, Gd), suggest that SC resides in the $CuO_2$ planes while MO resides in the $RuO_2$ sub-layers [3]. The Gd-sublattice orders with an anti-ferromagnetic structure at $T = 2.5$ K without affecting the SC state. The coexistence of SC and MO, when a ferromagnetic component is present, is always intriguing because of the detrimental effect an internal magnetic field can present to the formation of singlet Cooper pairs. A broad study was made of the coexistence of SC and MO in the intermetallics $RRh_4B_4$, $RMo_6S_8$ ($R$ = rare earth) [4], $ErNi_2B_2C$ [5], and others. Though the characteristics of SC in the oxides are different from those of the intermetallics, a large body of knowledge of this coexistence was developed with the study of the latter.

The notion of spatial separation between SC and MO has been one focal point in the study of the coexistence of these phenomena. It has been suggested that the coexistence of SC and MO in the Ru-1212-type compounds requires 1) small average magnetization at the $CuO_2$ SC planes, 2) the SC and MO subsystems are independent, and 3) coupling of the SC and MO layers is transverse to the layers, allowing for 3D-order of these ground states [6]. In addition, for SC and ferromagnetism (FM) to coexist in singlet pairing systems, the SC and FM order parameters need to develop spatial variations to accommodate each other [7, 8]. However this last requirement is only pertinent to singlet pair systems, and it would be less relevant to d-wave pairs as perhaps in Ru-1212. In light of the uncertainty regarding the exact pairing nature of the electrons in these systems, probing the coexistence is in order.

Long-range magnetic order in $RuSr_2GdCu_2O_8$ has been inferred from both magnetic susceptibility and muon spin relaxation (μSR) measurements [9, 10]. The μSR measurements suggested first that ferromagnetic ordering occurs at $T_m \approx 135$ K, due to the ordering of Ru moments, and anti-ferromagnetic ordering occurs at $T \approx 2.6$ K, due to the Gd moments. The values of the magnetic moments were reported to be $\mu(Ru) \approx 1.0 \ \mu_B$, and $\mu(Gd) \approx 7.4 \ \mu_B$. However, neutron diffraction studies showed that the magnetic structure due to the Ru moments, at $T_m \approx 135$ K, is that of a canted anti-FM (AFM), or weak-FM (wFM) [11, 12]. The magnitude of the magnetic moment along the c-axis is $\approx 1.18 \ \mu_B$, and the upper limit for the perpendicular

FM component is $\approx 0.1$ $\mu_B$. These neutron studies also confirm the anti-ferromagnetic ordering of the Gd moments at $T \approx 2.5$ K.

The occurrence of bulk superconductivity in $RuSr_2GdCu_2O_8$ has been determined by means of measurements of heat capacity and electrical resistivity ($\rho$) [2]. The Ru-1212 compounds show relatively broad SC transitions, compared to other oxide-based high-$T_c$ superconductors (HTS). The SC transition temperature, the width of the resistive transition $\Delta T_c$, and the normal state of resistivity all depend significantly on the synthesis [13]. The SC transition width $\Delta T_c$ in Ru-1212-Gd can be $\approx 8.5$ K in the best samples, to values 2-3 times higher. By contrast the width of transition to the SC state in the best quality $YBa_2Cu_3O_7$ (YBCO) samples is less than 1 K. The broad transition in the ruthenates has been attributed to motion of self-induced vortices [14], poor inter-granular connectivity [15], poor oxygenation, and the presence of impurities [1].

Most of the detailed studies of SC in the Ru-1212 compounds to date have been performed on polycrystalline specimens. The $\rho$ vs $T$ data for these specimens exhibit two distinct steps in the transition to the SC state. A useful paradigm for analyzing the bulk transport properties in these polycrystalline materials is a model in which the specimen is a 3D Josephson junction array (JJA) of weakly linked SC grains of Ru-1212. The two steps in $\rho$ vs $T$ at the onset of SC, at temperatures $T_c$ and $T_p$, are then attributed to transitions in the intra- and inter- granular regions, respectively. The distinction between intra- and inter- granular SC transition has been probed in detail by means of ac magnetic susceptibility $\chi_{ac}$ vs $T$ measurements in bulk polycrystalline and in powdered samples [16]. These measurements show that while the onset of intra-granular SC occurs at the same T for the bulk and powdered samples, inter-granular SC is quickly suppressed in the powdered the samples. This indicates that bulk SC can only be sustained by coupling through a JJA.

It has been suggested that intra-granular SC in Ru-1212 may rely on weak links as well. High-resolution electron microscopy (HRTEM) measurements indicate that the $RuO_2$ octahedra can be rotated about the *c*-axi*s* by 13° and -13°, giving rise to phase and anti-phase unit cells [17]. The Ru-1212 crystal is then a superstructure comprised of groupings of phased and anti-phased cells, leading to a divided structure, with subdomains of 50-200 Å. Thus the grains of Ru-1212 might themselves comprise a JJA, with nanoscale domains of SC.

The study of the coexistence of SC and MO leads to two challenging and related questions; first as to how SC can exist in the presence of wFM, and secondly how each of these phenomena can achieve sufficient coupling to its own planar species, while remaining decoupled from the other. Both SC and wFM are three-dimensional phenomena, and thus their coexistence requires that each couple through a spacer-layer of the other species.

First, the pair-breaking issue has been addressed by Fulde-Farrell-Larkin-Ovchinnikov (FFLO) type theories, wherein spatial variation of the order parameter of SC accommodates the existence of wFM [7, 8]. In this vein, it has been suggested that perhaps the SC order parameter changes phase by $\pi$ from SC layer to SC layer. This arrangement would lead to a node at the MO layer, thus decreasing the probability of pair breaking [6].

The issue of inter-layer magnetic coupling has been addressed, in part, by a calculation of the Hamiltonian ground states of a tri-layer (FM-SC-FM) superconducting ferromagnet [18, 19]. The calculation modeled two ferromagnetic Manganese Oxides sandwiching a superconducting Copper Oxide layer. The result was that the MO state oscillates between FM and AFM depending on the thickness of the SC layer.

The picture of coexistence of SC and MO in the Ru-1212 materials is one of spatial separation. The SC is confined to monolayer planes of Cu-O, and wFM is set on the planes of Ru-O octahedra, with weak coupling between the two. Thus the coexistence would be strongly dependent on both the arrangement of and the spacing between these two species of sub-layers. This spacer layer issue, along with the possibility of weak links in the intra- and inter- granular regions, warrants further study in which crystallographic structure/spacing is altered. Two common approaches to varying crystalline structure are chemical doping, and the application of hydrostatic pressure, the latter having the advantage of not introducing crystallographic disorder.

Lorenz et al. carried out a study of the pure Ru-1212-Gd compound, by means of $\rho$ vs $T$ measurements in hydrostatic pressures up to 2 GPa [20]. Their pressure study measured how $T_c$, $T_p$, and $T_m$ vary with pressure, and they found that all three transition temperatures increased with pressure. The rates of change of these transitions with pressure, the so-called pressure coefficients, were determined to be: $dT_c/dP \approx 1.0$ K/GPa, $dT_p/dP \approx 1.8$ K/GPa, and $dT_m/dP \approx 6.7$ K/GPa. Thus the pressure enhancement of $T_m$ is $\approx 6$ times that of $T_c$. These authors posit that the stronger enhancement of magnetism with pressure, relative to SC, suggests a strong competition between SC and MO in the material. This notion is supported by contrasting the pressure behavior of Ru-1212 with YBCO, the canonical under-doped cuprate superconductor (UCSC) in which $dT_c/dP \sim 3\text{-}4$ K/GPa [21, 22]. The salient difference is that while Ru-1212 exhibits a coexistence of SC and MO, YBCO does not. Absent the competition between SC and MO in Ru-1212, the values of $dT_c/dP$ in YBCO are 3-4 times higher than in Ru-1212. The argument has some credence because Ru-1212 can be regarded as a UCSC because the charge carrier density is $n_h \approx 0.1$ holes/Cu, which is typical for these superconductors [17], and because its crystal structure is derived from YBCO's, by the replacement of the Cu-O charge reservoirs with $RuO_6$ octahedra. In contrast to Ru-1212, the effect of hydrostatic pressure on the related $RuSr_2(Sm_{1.4}Ce_{0.6})Cu_2O_{10}$ compound is to increase $T_c$ at the rate of 4.7 K/GPa, while $T_m$ drops at the rate of –12.9 K/GPa [23], i.e., a drop in $T_m$ permits an increase rate $T_c$ with pressure even higher than in YBCO, supporting again the notion of competition between SC and magnetism. The coexistence of SC and magnetic order in the Ru-1212 compounds has been further probed by dilution of the Ru sublattice, as for example by partial substitutions with Ti [24], V, Nb [25], Rh [24, 26], Ta [27], and Ir [28]. A noteworthy pressure result illustrative of the competition between SC and magnetic order was obtained by Yamada et al. [25]; Ru-1212 specimens with 10% substitution of V and Nb at the Ru site show an increase and a decrease in $T_m$, respectively. The effect of pressure is to depress $T_c$ on the material with increased $T_m$ (V), and to increase $T_c$ in the material with deppressed $T_m$ (Nb).

In order to probe further the coexistence of SC and MO in these granular hybrid rutheno-cuprates, we carried out measurements of the pressure dependence of $\rho$ vs $T$ in $Ru_{1-x}Rh_xSr_2GdCu_2O_8$ compounds ($x = 0$-$0.15$), in pressures up to 2 GPa. The partial substitution of Rh for Rh can be accomplished up to $x \approx 0.25$ without noticeable changes in phase purity, and crystal structure, including lattice parameters [24, 26]. The partial substitution of Rh for Ru causes a fast drop in both $T_m$, and $T_c$, and it promotes granularity. The detrimental effect of disorder on the charge transfer between the adjacent Ru(Rh)-O and Cu-O layers is a strong impediment to SC, and it makes it difficult to analyze the effect of doping on $T_c$, and $T_m$, using charge arguments. The effect of pressure on the intra- and inter- granular weak-links was probed by studying the effect of the excitation current on the resistive transition to the SC state.

**Experimental Details**

The samples of the $Ru_{1-x}Rh_xSr_2GdCu_2O_8$ compounds with x = 0, 0.05, 0.10, and 0.15.were synthesized by solid-state reaction. First a $Sr_2(Ru,Rh)GdO_6$ (Sr-2116) precursor phase was made from a stoichiometric mixture of high purity Ru, Rh, $Gd_2O_3$, and $SrCO_3$. This precursor was then reacted with CuO to form $RuSr_2GdCu_2O_8$, as previously reported [26, 29]. The crystal structure of the samples was verified using a Philips Analytical X'Pert MPD Pro Theta/Theta X-ray diffractometer (XRD); the lattice parameters, and the presence of impurities were determined by Rietveld analysis using the Philips X'pert software. The XRD results showed that all samples up to $x = 0.25$ were nearly single-phase, although minute amounts of $Sr(Ru,Rh)O_3$ and Sr-2116 impurity phases could be detected. The lattice parameters didn't change noticeably with Rh concentration, and they matched those published in the literature [24].

The measurements of electrical $\rho$ vs $T$ under pressure were carried out in a piston-cylinder self-clamping Be-Cu cell, with a non-magnetic CrNiAl alloy hardened core, manufactured at the High Pressure Physics Institute in Moscow, Russia [30].  The outer diameter of the cell is 30 mm, the pressure cavity is cylindrical, about 30 mm long, and 6.2 mm in diameter.  Electrical leads for the four-wire measurements of electrical resistivity of the sample and a manganin manometer were routed into the cell via a stycast sealed feedthrough made with the CrNiAl alloy.  Four Pt leads were attached to the sample with silver epoxy, and the sample was glued to a Kapton tape platform resting on top of the feedthrough wires.  The typical contact resistance between the Pt leads and sample were 1-2 Ω.  The sample was then inserted into a teflon capsule (inner diameter = 5.5 mm, length = 20 mm) filled with Fluorinert-FC75, which was used as the pressure transmitting medium.  Force was applied with a hydraulic press at room temperature (RT) through an assembly that included a tungsten carbide piston and an anti-extrusion disc in direct contact with the bottom of the teflon capsule. The RT pressure ($P_{300K}$) was determined by monitoring the change in resistance of a coil of manganin wire [30], and the pressure was locked in with a Be-Cu lock-nut.  Since the pressure in this type of cell is typically reduced as temperature is lowered [31], the pressure values below 300 K needed to be estimated.  First we used a previous cross-calibration measurements with this cell in which both manganing and Sn manometers were simultaneously in the pressure chamber; the pressure at 300 K was determined with the manganin manometer, and the pressure below 4 K was determined from the superconducting transition temperature in Sn [32].  We then assumed a linear drop in pressure from 300 to 90 K [31], and assumed that below 90 K the pressure remained the same as the pressure yielded by the Sn manometer.  Additionally, by monitoring the change in resistance of the manganin manometer with temperature, and assuming a negligible change in $R(P)$ with temperature, we confirmed the manganin/Sn estimate within a few percent.

Cooling was accomplished by attaching the pressure cell to the cold finger of a close-cycle He refrigerator.  Thermally conducting silver grease (Heataway 641) was applied to the surface between the cold-finger and the cell, and to the threads of the lock-nuts, in order to assure good thermal coupling.  The temperature of the cell was monitored with a Si-diode placed on the outer surface of the cell, at a thermal distance from the cold-finger close to the sample's.  Each sample was loaded in the pressure cell, and a no-load $\rho$ vs $T$ measurement was made in the T-range from 10-300 K, while the sample was in contact with the pressure-transmitting medium.  The sample's resistance was measured with a low-power ac 4-wire resistance bridge (Linear Research LR-400) operating at 16 Hz.  The pressure at RT was then incremented in steps of ≈ 0.5 GPa up to ≈ 2 GPa, and a new $\rho$ vs $T$ measurement was carried out.  In light of the difference in contraction of the pressure transmitting medium and the body of the cell, the no-load is not a truly no-pressure measurements, and the corresponding $\rho$ vs $T$ data is shown just as a reference. For each set of $\rho$ vs $T$ measurements the cell was first cooled from 300 to 10 K at a rate of $\approx 2.5$ K/min. While the cool-down data was monitored closely in order to identify the transition temperatures, the lag in temperature between the Si-diode T-sensor and the sample was large and inadequate for our analysis. The reliable $\rho$ vs $T$ data from 10 to 50 K was collected while the refrigerator was on, and a heater provided just enough heat to raise the temperature of the cell at the rate $\leq 0.5$ K/min. For the measurements above 50 K, both the refrigerator and the heater were turned off, and the cell was let warm up naturally at a rate of $\leq 0.5$ K/min. The lag in temperature between the sample and temperature sensor was estimated to be below 0.2 K.

In order to probe the effect of pressure on the weak links of these granular superconductors, we carried out measurements of $\rho$ vs $T$ at $P_{300K} \approx 0$ (no load), and 1.5 GPa, with varied excitation currents ($I_{ex}$) in the 30 µA to 10 mA range. These measurements were carried out on the $x = 0$ and 0.10 Rh-doped samples.

**Experimental Results**

The effect of Rh doping on the temperature dependence of electrical resistivity is shown in Fig. 1. These data were collected with the samples mounted inside of the pressure cell, with the Teflon capsule filled with the pressure-transmitting medium, but no load was applied. The undoped sample shows metallic behavior for $T > 70$ K. The behavior of $\rho$ vs $T$ in this 70-300 K T-interval is linear, with $d\rho/dT \approx 0.4$ mΩ-m/K. The value of $\rho$ starts to decrease below $\approx 55$ K, at the rate $d\rho/dT \approx 8.9$ mΩ-m/K, which marks the onset of intra-granular SC. At 50 K the resistivity starts decreasing at a slightly faster rate, $d\rho/dT \approx 9.5$ mΩ-m/K, indicating the onset of inter-granular SC. The resistivity finally reaches zero around 40 K. The onset of magnetic ordering is manifest by a noticeable feature in $\rho$ near $\approx 150$ K.

With partial substitution of Rh for Ru, the values of the slopes $d\rho/dT$ for T > 70 K are reduced. The slopes $d\rho/dT$ in the intra- and inter- granular regions becoming more differentiated, as seen in the inset of Fig. 1 for the 5% Rh sample, suggesting that Rh-doping also promotes granular behavior. In addition, the feature in $\rho$ vs $T$ near $T_m$ becomes more apparent with Rh-doping. The three transition temperatures $T_c$, $T_p$, and $T_m$ are found to decrease with doping; with $dT_c/d(\%Rh) = -2.60$ K/%Rh, $dT_p/d(\%Rh) = -1.40$ K/%Rh, and $dT_m/d(\%Rh) = -0.20$ K/%Rh.

The effect of the hydrostatic pressure in all compositions is to raise the values of $T_c$, $T_p$, and $T_m$. An example of the evolution of the normalized $\rho$ vs $T$ curve with pressure is shown in Fig. 2a, for the 5% Rh-doped sample. The pressure values cited are the nominal values at room temperature. Two trends are evident with the application of pressure: 1) the decrease in normal state resistivity, and 2) the increase in the intra- and inter- granular SC transition temperatures. Neglecting the no-load data, the decrease in the resistivity of the normal phase is approximately linear, as shown in Fig. 2b for $T = 50$ K.

The values of $T_c$, $T_p$, and $T_m$ were determined both from fitting $d\rho/dT$ to a double Gaussian ($x = 0$, and 0.05 only) [20], as shown in Fig. 3a, and from linear extrapolations (all compositions), as shown in Fig. 3b. The values of $T_m$ were determined from the peak in $d\rho/dT$ at the onset of magnetic ordering, as shown in the inset of Fig. 3a for the undoped sample. Since the onset of inter-granular SC for the $x = 0.10$ and 0.15 samples falls below 10K, the lowest $T$ of our apparatus, only the onset of intra-granular SC can be seen clearly and the double Gaussian determination of $T_c$ and $T_p$ could not be made. Because of the pressure effect, the inter-granular SC transition of the $x = 0.10$ samples becomes visible at higher pressures, and the intra-granular SC transition of the $x = 0.15$ sample appears more complete (data not shown).

The value of $T_m$ increases almost linearly with pressure for all compositions studied, as shown in Fig. 4, and the value of $dT_m/dP$ varied in the range from 3.7 to 7.9 K/GPa. The value of $dT_m/dP$ first decreases for small Rh concentration and then starts to increase as more Rh is added. The values of $T_c$ and $T_p$ also increase in a nearly linear fashion with pressure, as shown in Figs. 5a and 5b, and the values of $dT_c/dP$ and $dT_p/dP$ are in the 1.1-1.8 and 1.5-3.1 K/GPa ranges, respectively. Similarly to $T_m$, the values of $dT_c/dP$ and $dT_p/dP$ first decrease for small amounts of Rh, then increase again as more Rh is added, as shown in Fig. 6.

In order to probe the granular characteristic of the SC in these Ru-1212 materials, we carried out measurements of the effect of the excitation current on the behavior of $\rho$ vs $T$. The $I_{ex}$ for the 4-wire measurement was varied in the 30 µA to 10 mA range, at the pressures $P_{300K} = 0$, and 1.5 GPa. The effect of high $I_{ex}$ in the $\rho$ vs $T$ measurement is the broadening of the SC transition for both the undoped (Fig. 7a), and Rh-doped materials (data not shown). While the effect of $I_{ex}$ is quite noticeable for $I_{ex} > 1$ mA in $P_{300K} = 0$, it is drastically suppressed under pressure (Fig. 7b), suggesting better inter-granular connectivity. It should be noted that the broadening in $\rho$ vs $T$ near $T_c$ is somewhat more noticeable for samples placed in the pressure cell, where they are in contact with the pressure-transmitting medium, than when they are not.

Plots of $\Delta T_c$ and $\Delta T_p$ vs $I_{ex}$, where the $\Delta$'s represent the changes in apparent $T_c$ and $T_p$ due to $I_{ex}$ larger than 30-100 µA, are shown in Fig. 8. Both the 0 and 10% Rh-doped samples show a significant reduction of the $I_{ex}$ effect with pressure. In the pure sample, the value of $dT_c/dI_{ex}$ changes drastically from -1.0 K/mA to -4.5x10$^{-2}$ K/mA as pressure is increase from zero to 1.5 GPa.

**Conclusions**

We carried out measurements of electrical resistivity versus temperature under hydrostatic pressure up to 2 GPa in Ru$_{1-x}$Rh$_x$Sr$_2$GdCu$_2$O$_8$ compounds with $x$ = 0, 0.05, 0.10, and 0.15 in order to probe both the coexistence of superconductivity and magnetism, and the granular nature of the superconductivity. The partial substitution of Rh for Ru lowers $T_m$, $T_c$, and $T_p$, and it makes the granular character of the bulk superconductivity much more noticeable. The application of pressure to these materials increases the transition temperatures $T_m$, $T_c$, and $T_p$. The values of $dT_m/dP$ are 3-4 times higher than $dT_c/dP$ for all compositions. The change in $T_c$ with pressure ranged from 1.1 to 1.7 K/GPa in the compositions studied, values that are 2-3 times lower than observed in YBCO. In light of the concomitant increase in $T_m$ with pressure, it is tempting to ascribe the lower $dT_c/dP$ to the competition between SC and MO, as suggested in Ref. 20. This argument is strengthened by the finding the effect of pressure on the related material RuSr$_2$(Sm$_{1.4}$Ce$_{0.6}$)Cu$_2$O$_{10}$ is to raise $T_c$ at the much higher rate of 4.7 K/GPa, while $T_m$ drops at the rate of –12.9 K/GPa [23].

Excitation currents in excess of 1 mA ($J \approx 0.1$ A/cm$^2$) start broadening the SC transitions and shifting them to lower temperatures. Both these effects are suppressed by the application of hydrostatic pressure, suggesting the possibility that the weak links are due to poor mechanical contact between the grains.

The problems of whether the superconductivity and magnetism are competing in these ruthenates, and whether the weak links are exclusively inter-granular or can be intra-granular as well are still open questions. Further work on single-crystals is in order.

*Acknowledgements*

The support from NSF grant No. DMR-0306165 for the work at SDSU is gratefully acknowledged.

**Figure Captions**

Figure 1. Plot of $\rho/\rho_{300K}$ vs $T$ for $Ru_{1-x}Rh_xSr_2GdCu_2O_8$ with $x = 0$, 0.05, 0.10, and 0.15. These measurements were taken with the samples loaded in the pressure cell, and in contact with the pressure-transmitting medium; however, no load was applied. The inset shows the differentiation between the intra- and inter- granular SC transitions for the 5% Rh sample.

Figure 2. (a) Normalized $\rho/\rho_{300K}$ vs $T$ for $Ru_{0.95}Rh_{0.05}Sr_2GdCu_2O_8$. The pressures shown are the nominal values at room temperature. (b) Normalized $\rho_{50K}/\rho_{300K}$ vs $P$ in the same sample. The pressure values shown were corrected to reflect the drop in pressure of the cell for $T < 300$ K. The solid line is a linear fit ignoring the no-load data.

Figure 3. (a) Determination of $T_c$ and $T_p$ in $RuSr_2GdCu_2O_8$ by two-gauss fit. The inset shows the determination of $T_m$ from the peak in $d\rho/dT$. (b) Normalized $\rho/\rho_{300K}$ vs $T$ for $Ru_{0.90}Rh_{0.10}Sr_2GdCu_2O_8$. Inset shows the determination of $T_c$ and $T_p$ from the interception of the extrapolated straight lines.

Figure 4. Pressure dependence of $T_m$ for $Ru_{1-x}Rh_xSr_2GdCu_2O_8$ with $x = 0$, 0.05, 0.10, and 0.15, showing the increase in $T_m$ with $P$. Solid lines are linear fits including the "high-load" data only; the dotted lines are extrapolations of the linear fits into the "no-load" regime.

Figure 5. Pressure dependence of (a) $T_c$ and (b) $T_p$ for $Ru_{1-x}Rh_xSr_2GdCu_2O_8$ with x = 0, 0.05, 0.10, and 0.15. Solid lines are linear fits including the "high-load" data only; the dotted lines are extrapolations to low $P$. The values $T_c$ and $T_p$ extracted from double-Gauss (DG) for $x = 0$ are shown for comparison. All slopes are listed in legend.

Figure 6. Dependence of the pressure coefficients $dT_c/dP$, $dT_p/dP$, and $dT_m/dP$ with Rh-doping. The solid lines are guides to the eye.

Figure 7. Normalized $\rho/\rho_{300K}$ vs $T$ for $RuSr_2GdCu_2O_8$ for excitation currents between 30 µA and 10 mA. The nominal pressures at 300K are (a) $P_{300K} = 0$ (no load); and (b) $P_{300K} = 1.5$ GPa. These data show a significant decrease in current dependence with the addition of pressure.

Figure 8. Semi-log plot of $\Delta T_c$ and $\Delta T_p$ as a function of excitation current, extracted from the $\rho$ vs $T$ data for (a) $x = 0$, and (b) $x = 0.1$. The solid lines are linear fits to the $P = 0$ data, and the dashed lines are linear fits to the $P = 1.5$ GPa data.

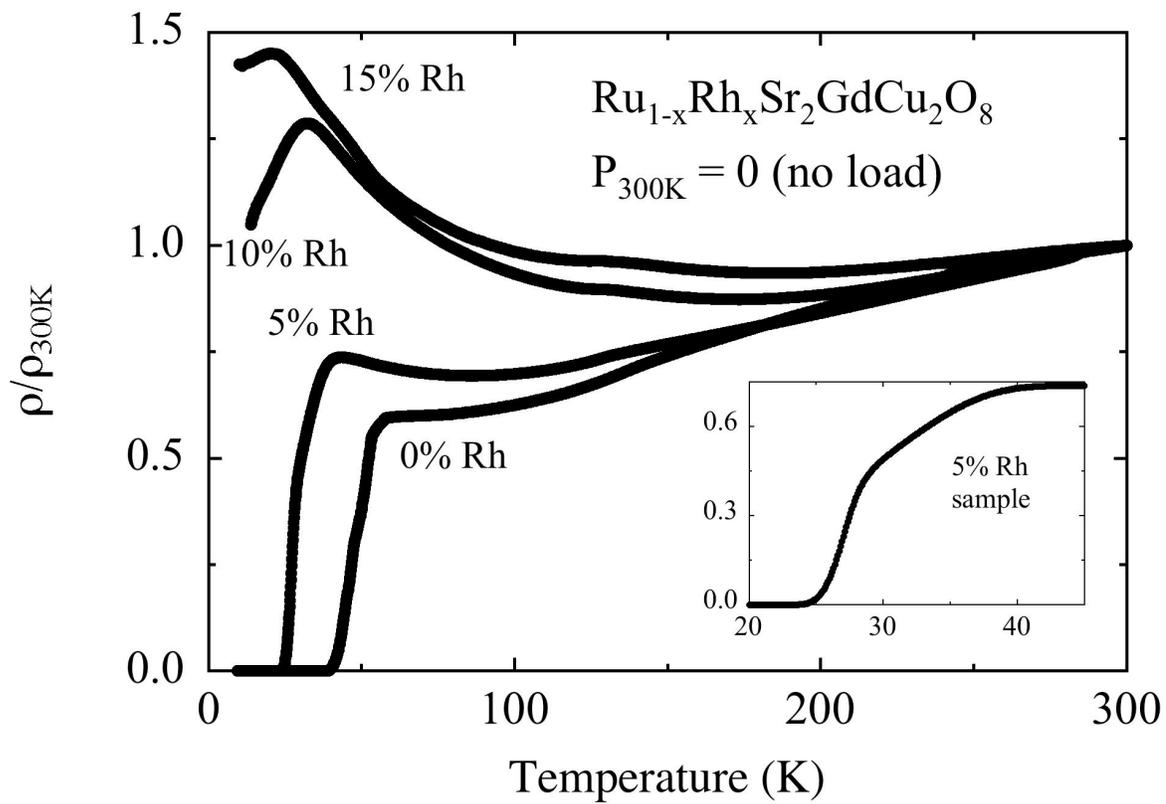

Figure 1

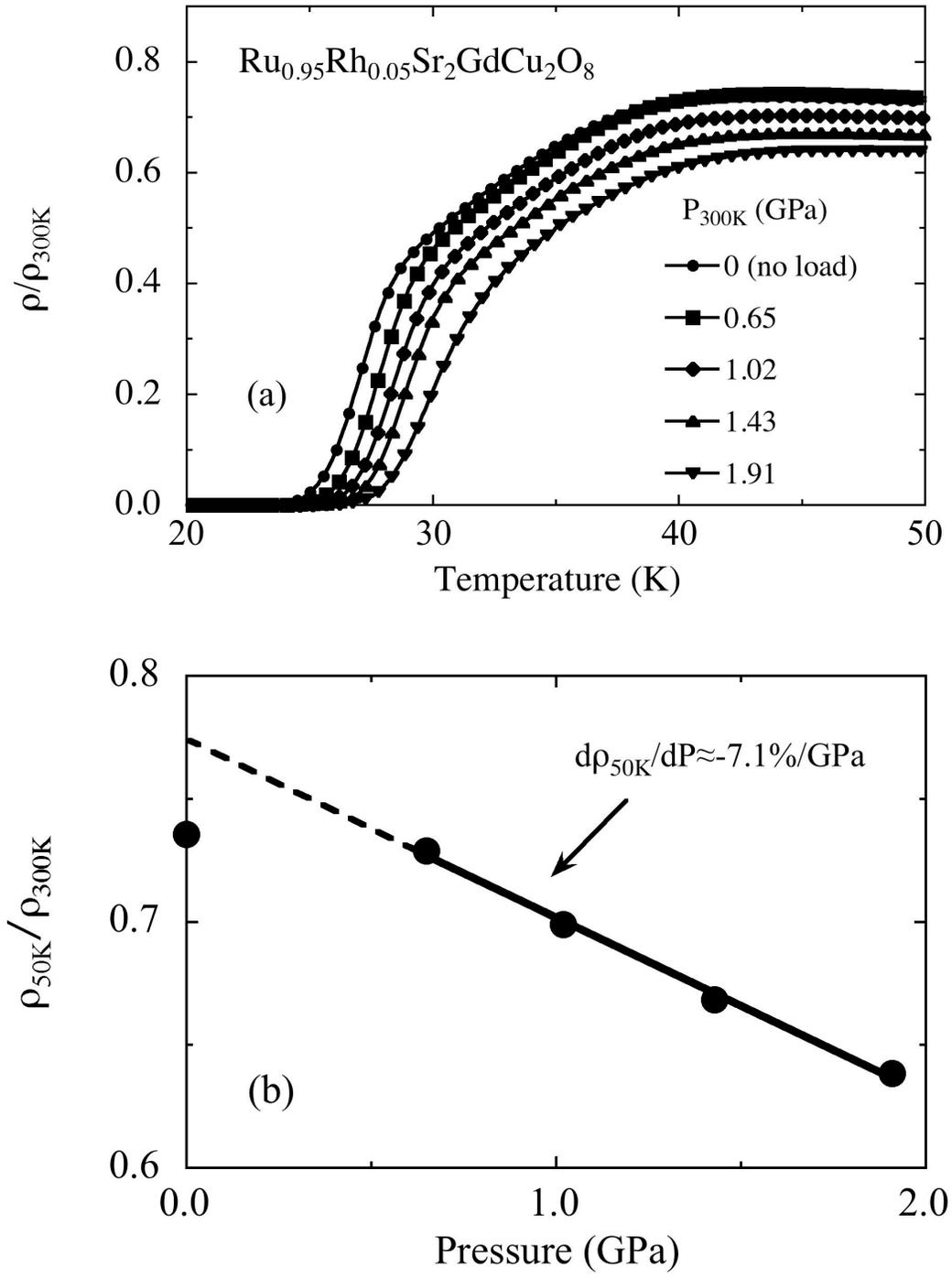

Figure 2

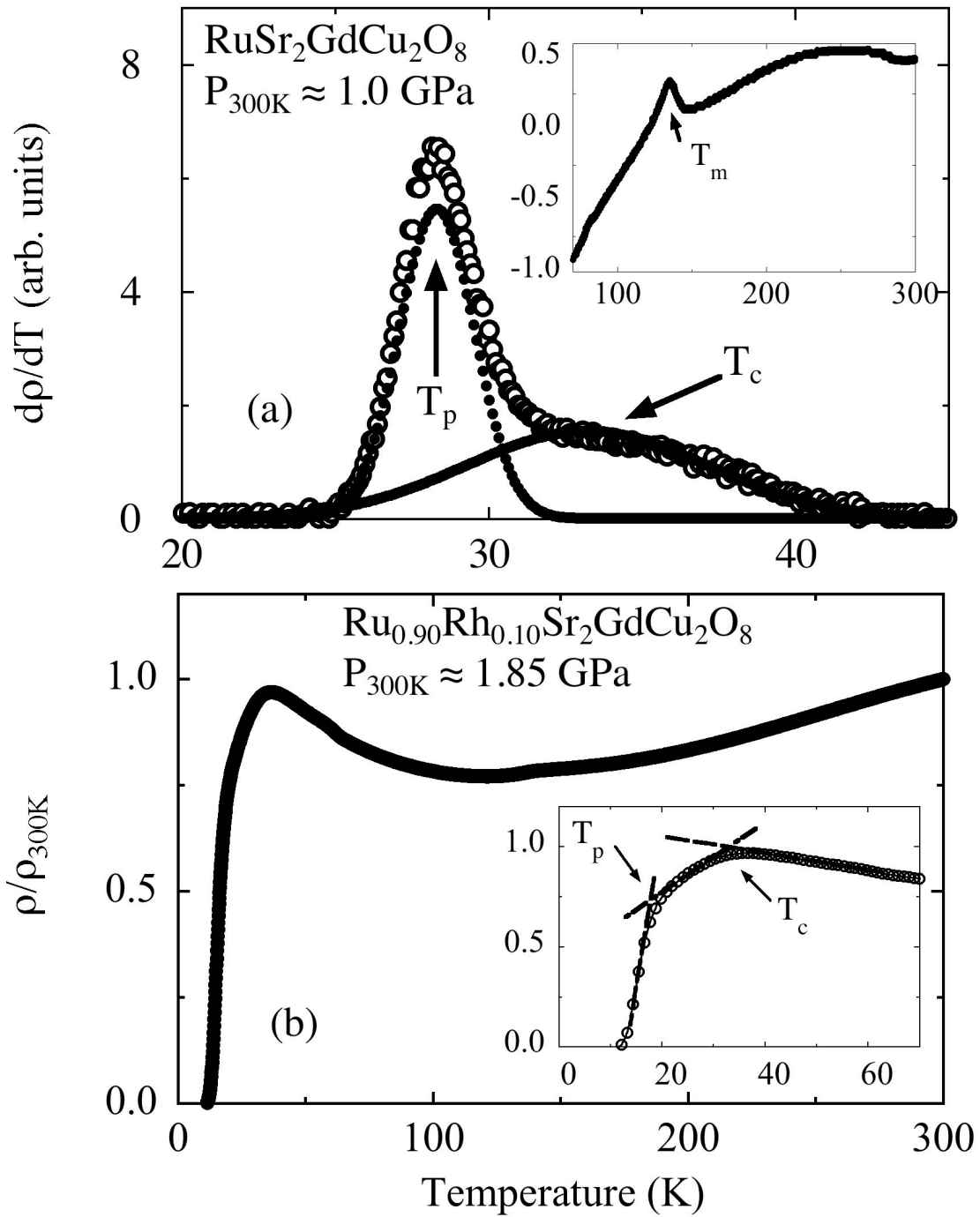

Figure 3

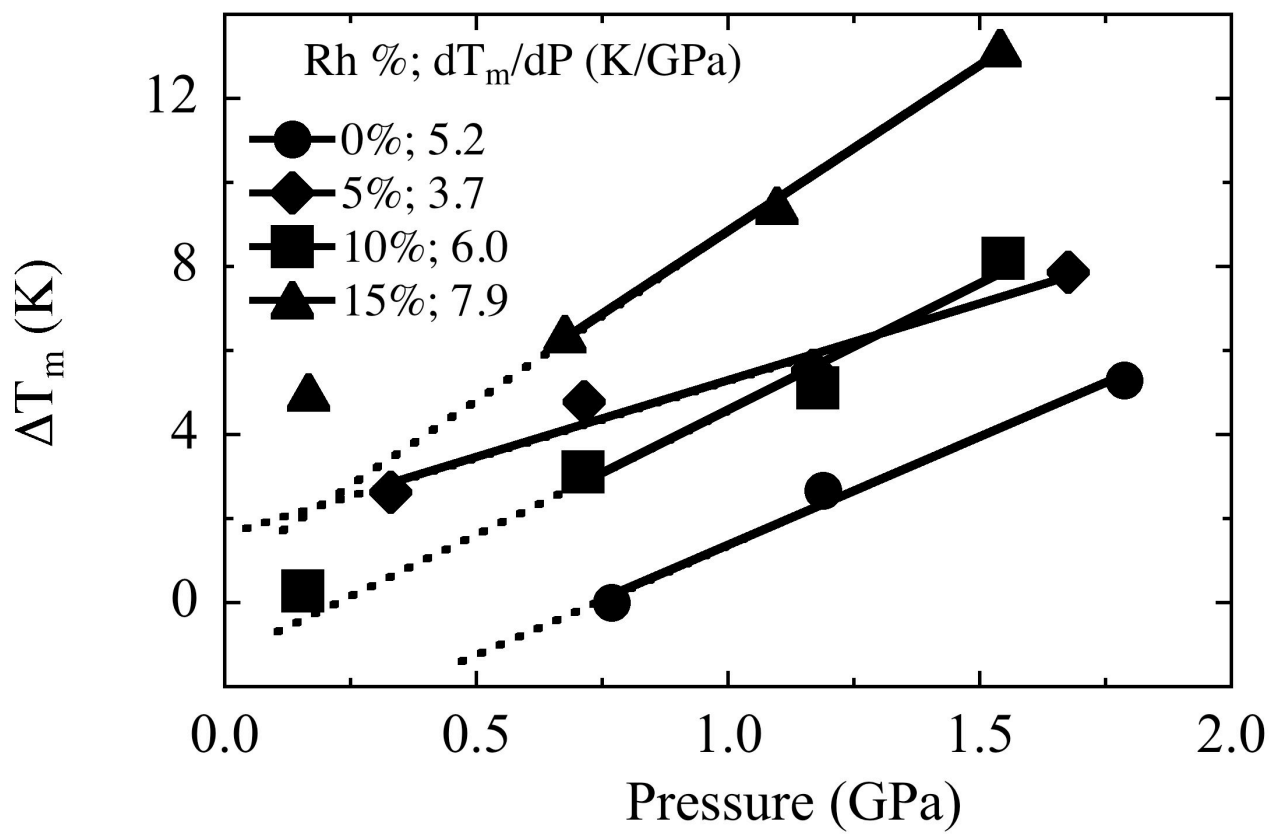

Figure 4

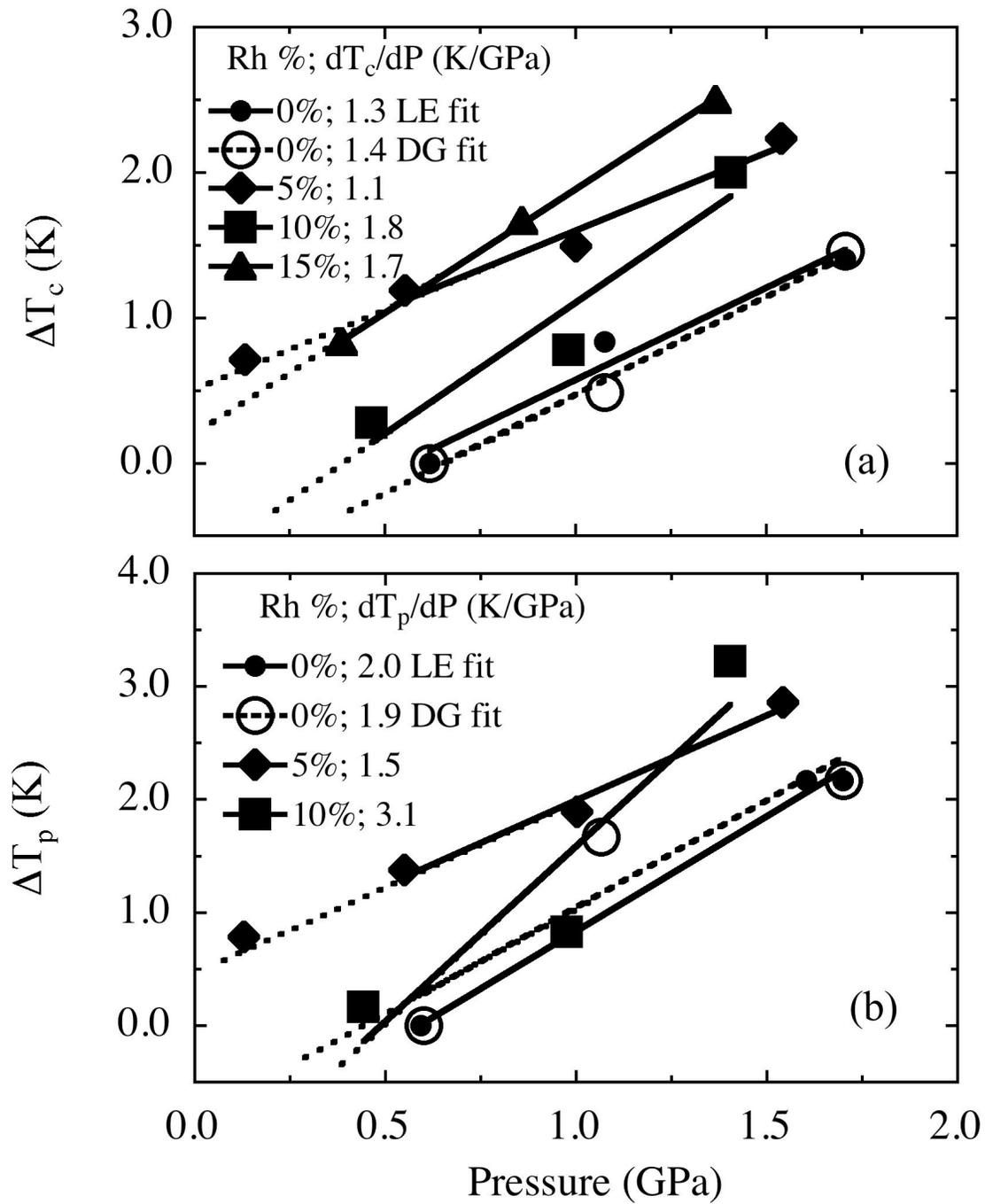

Figure 5

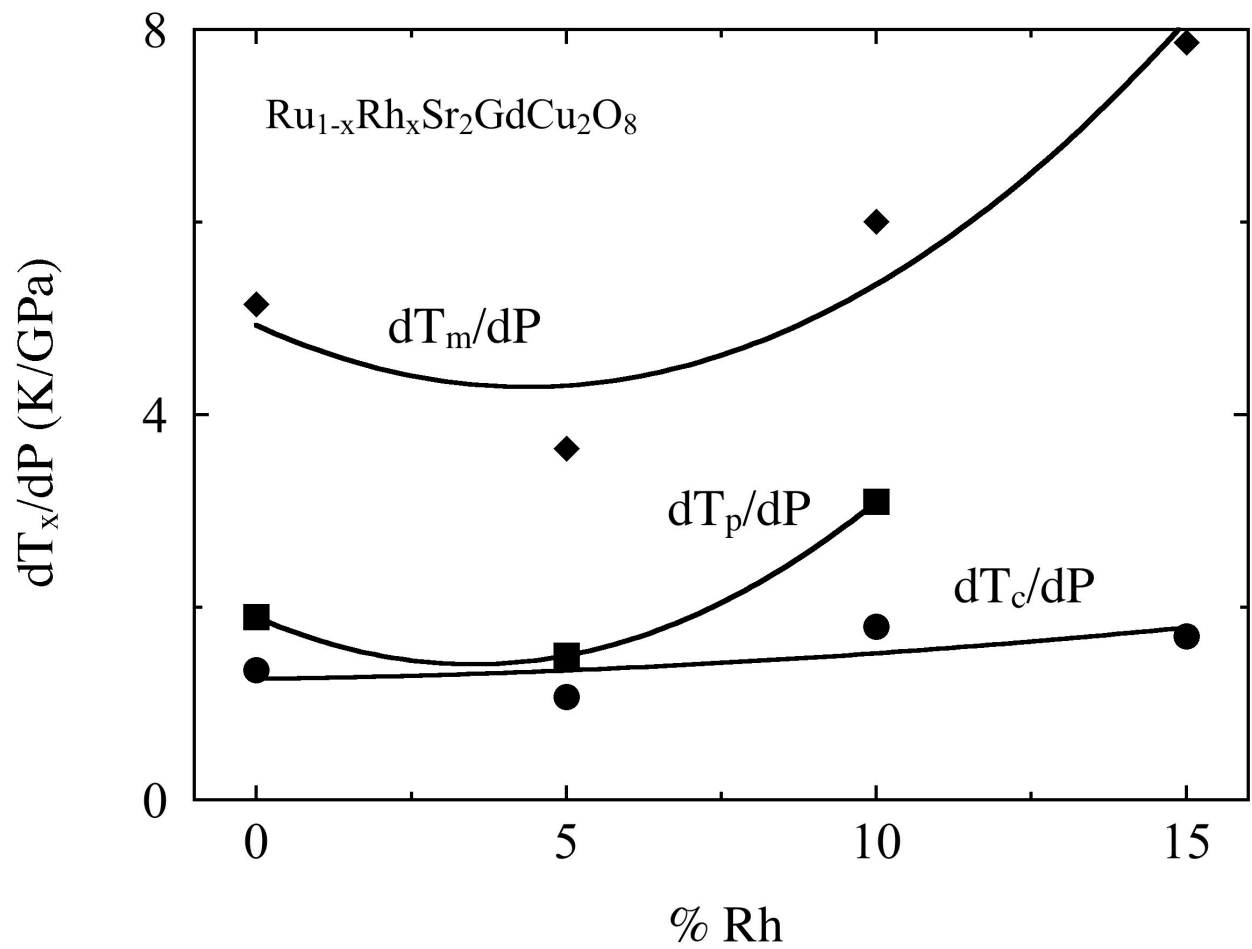

Figure 6

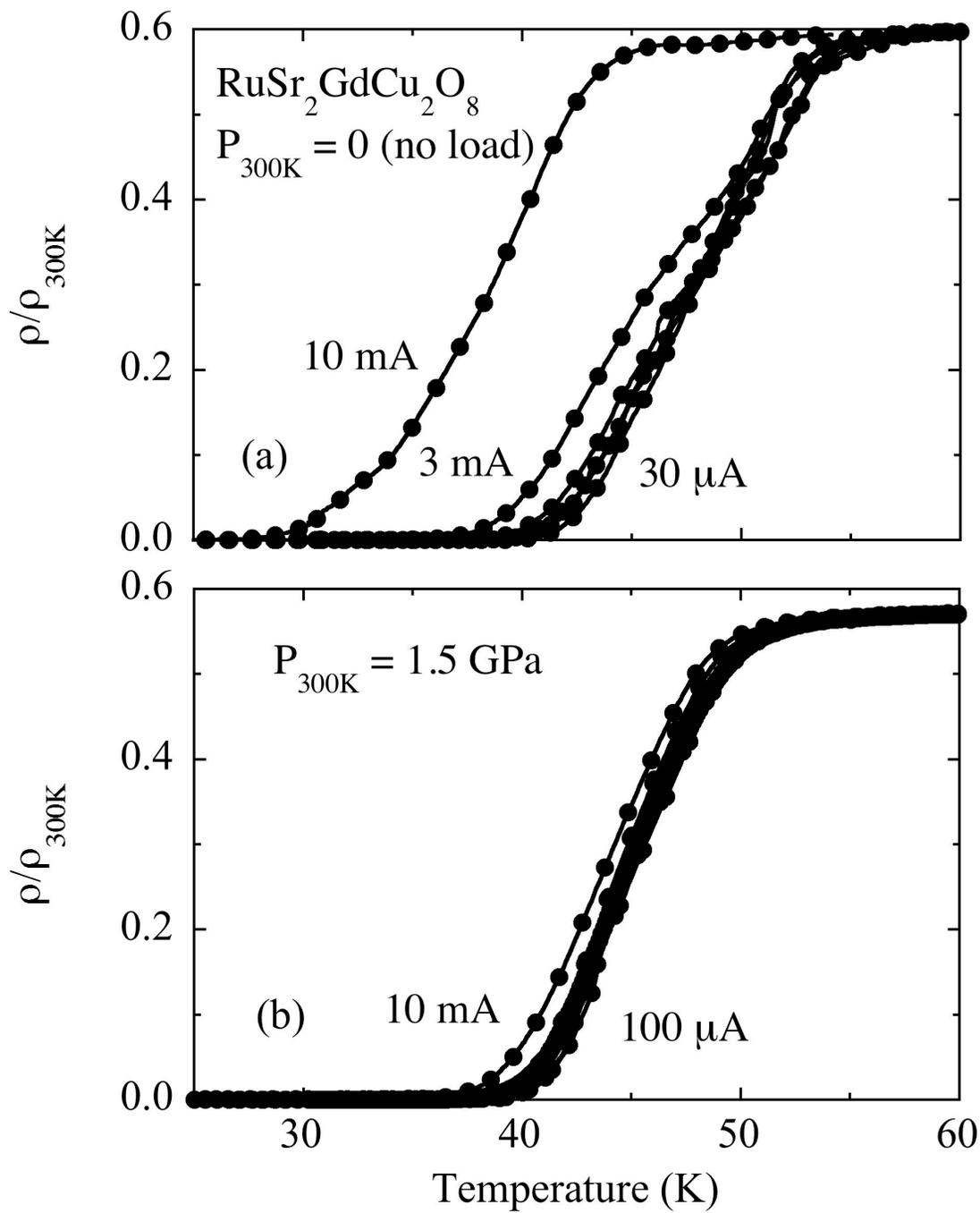

Figure 7

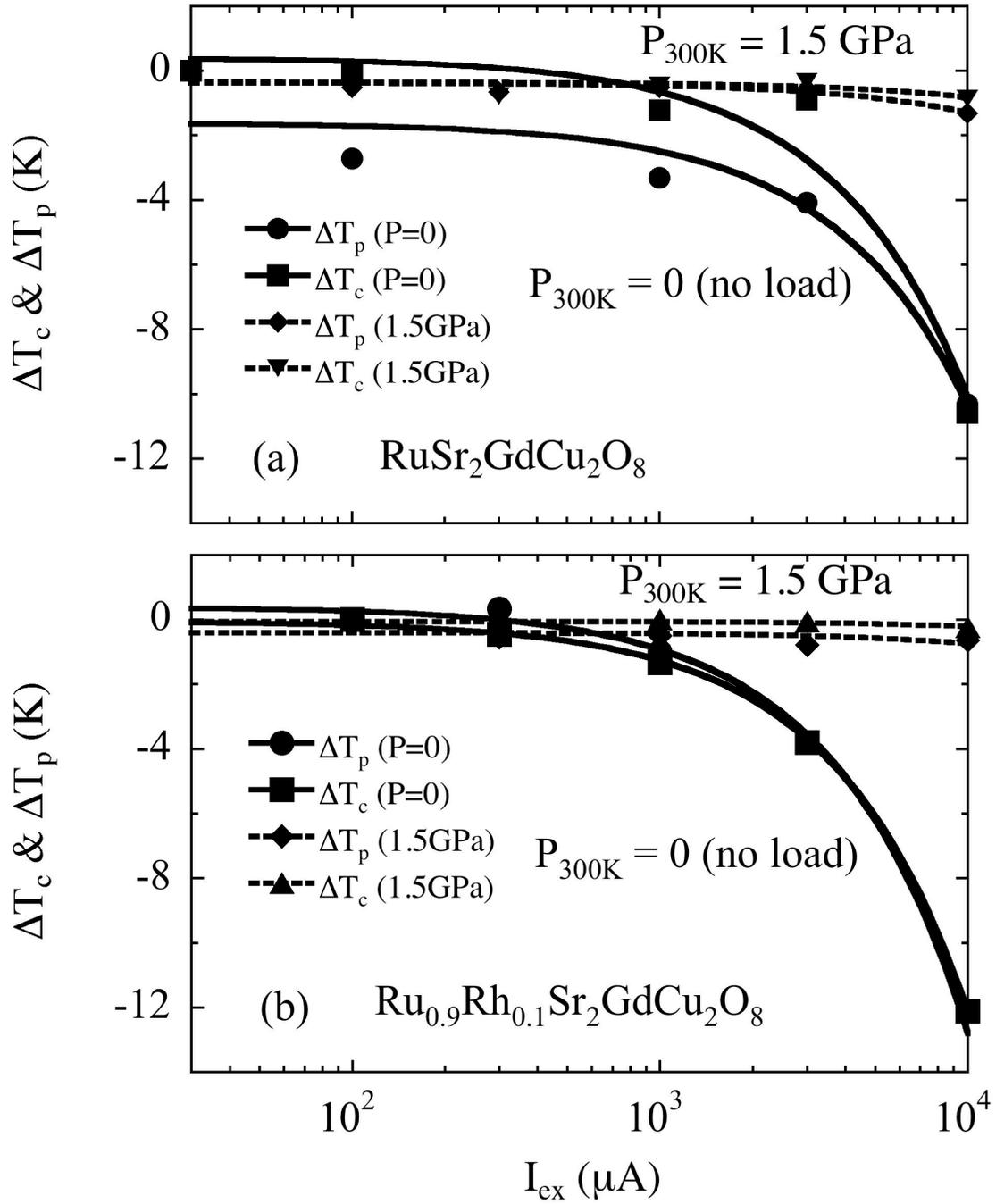

Figure 8

# References

[1]     L. Bauernfeind, W. Widder, and H. F. Braun, *Physica C* **254**, 151 (1995).

[2]     J. L. Tallon, J. W. Loram, G. V. M. Williams, and C. Bernhard, *Phys. Rev. B* **61**, 6471 (2000).

[3]     I. Felner, U. Asaf, Y. Levi, and O. Millo, *Phys. Rev. B* **55**, R3374 (1997).

[4]     M. B. Maple, and Ø. Fischer, *(eds.) Superconductivity in Ternary Compound, Topics in Current Physics, Vols. 32 & 34* (Springer-Verlag, Berlin, 1982).

[5]     P. C. Canfield, S. L. Bud'ko, and B. K. Cho, *Physica C* **262**, 249 (1996).

[6]     W. E. Pickett, R. Weht, and A. B. Shick, *Phys. Rev. Lett.* **83**, 3713 (1999).

[7]     P. Fulde, and R. A. Ferrell, *Phys. Rev.* **135**, 550 (1964).

[8]     A. I. Larkin, and Y. N. Ovchinnikov, *Sov. Phys. JETP* **20**, 762 (1965).

[9]     I. Felner, U. Asaf, S. Reich, and Y. Tsabba, *Physica C* **311**, 163 (1999).

[10]    C. Bernhard, J. L. Tallon, C. Niedermayer, T. Blasius, A. Golnik, E. Brücher, R. K. Kremer, D. R. Noakes, C. E. Stronach, and E. J. Ansaldo, *Phys. Rev. B* **59**, 14099 (1999).

[11]    J. W. Lynn, Keimer, Ulrich, C. Bernhard, and J. L. Tallon, *Phys. Rev. B* **61**, 14964 (2000).

[12]    O. Chmaissem, J. D. Jorgensen, H. Shaked, P. Dollar, and J. L. Tallon, *Phys. Rev. B* **61**, 6401 (2000).

[13]    C. Bernhard, J. L. Tallon, E. Brücher, and R. K. Kremer, *Phys. Rev. B* **61**, 14960 (2000).

[14]    Y. Tokunaga, H. Kotegawa, K. Ishida, Y. Kitaoka, H. Takagiwa, and J. Akimitsu, *Phys. Rev. Lett.* **86**, 5767 (2001).

[15]    R. L. Meng, B. Lorenz, Y. S. Wang, J. Cmaidalka, Y. Y. Xue, and C. W. Chu, *Physica C* **353**, 195 (2001).

[16]    B. Lorenz, Y. Y. Xue, R. L. Meng, and C. W. Chu, *Phys. Rev. B* **65**, 174503 (2002).

[17]    A. C. McLaughlin, W. Zhou, J. P. Attfield, A. N. Fitch, and J. L. Tallon, *Phys. Rev. B* **60**, 7512 (1999).

[18]    C. A. R. Sá de Melo, *Phys. Rev. B* **62**, 12303 (2000).

[19]    C. A. R. Sá de Melo, *Physica C* **387**, 17 (2003).

[20]    B. Lorenz, R. L. Meng, Y. Y. Xue, and C. W. Chu, *Physica C* **383**, 337 (2003).

[21]    W. H. Fietz, R. Quenzel, H. A. Ludwig, K. Grube, S. I. Schlachter, F. W. Hornung, T. Wolf, A. Erb, M. Klaser, and G. Muller-Vogt, *Physica C* **270**, 258 (1996).

[22]    Q. Xiong, Y. Y. Xue, J. W. Chu, Y. Y. Sun, Y. Q. Wang, P. H. Hor, and C. W. Chu, *Phys. Rev. B* **47**, 11337 (1993).

[23] G. Oomi, H. Fuminori, M. Ohashi, T. Eto, D. P. Hai, S. Kamisawa, M. Watanabe, and K. Kadowaki, *Physica B* **312-313**, 88 (2002).

[24] A. Hassen, J. Hemberger, A. Loidl, and A. Krimmel, *Physica C* **400**, 71 (2003).

[25] Y. Yamada, H. Hamada, M. Shimada, S. Kubo, and A. Matsushita, *J. Magn. Mag. Mat.* **272-276**, E173 (2004).

[26] M. S. Torikachvili, M. Steiger, L. Harding, D. Bird, N. Dilley, S. Gomez, J. O'Brien, and R. F. Jardim, in *Proc. Int. Conf. Low Temp. Phys. LT24* (AIP, Orlando, FL, 2005), Vol. 850, p. 679.

[27] X. H. Chen, Z. Sun, K. Q. Wang, Y. M. Xiong, H. S. Yang, H. H. Wen, Y. M. Ni, and Z. X. Zhao, *J. of Phys. (Cond. Matt.)* **12**, 10561 (2000).

[28] M. S. Torikachvili, I. Bossi, J. R. O'Brien, F. C. Fonseca, R. Muccillo, and R. F. Jardim, *Physica C* **408-410**, 195 (2004).

[29] T. P. Papageorgiou, T. Herrmannsdoorfer, R. Dinnebier, T. Mai, T. Ernst, M. Wunschel, and H. F. Braun, *Physica C* **377**, 383 (2002).

[30] M. I. Eremets, *High pressure experimental methods* (Oxford University Press, 1996).

[31] J. D. Thompson, *Rev. Sci. Instrum.* **55**, 231 (1984).

[32] T. F. Smith, C. W. Chu, and M. B. Maple, *Cryogenics* **9**, 53 (1969).